\documentclass[letterpaper, 10 pt, conference]{ieeeconf}  

\IEEEoverridecommandlockouts                              

\usepackage{graphicx}
\graphicspath{{./images/}}
\usepackage{amsmath}
\usepackage{amssymb}
\usepackage{booktabs}
\newcommand{\ra}[1]{\renewcommand{\arraystretch}{#1}}
\usepackage{multirow}
\usepackage{siunitx}
\usepackage{balance}

\title{\LARGE \bf
Modeling Electromagnetic Navigation Systems for Medical Applications using Random Forests and Artificial Neural Networks
}

\author{Ruoxi Yu$^{1}$, Samuel L. Charreyron$^{2}$, Quentin Boehler$^{2}$, Cameron Weibel$^{2}$, \\ Carmen C. Y. Poon$^{1}$ and Bradley J. Nelson$^{2}$
\thanks{$^{1}$R. Yu and C.C.Y. Poon are with the Division of Biomedical Engineering Research, Department of Surgery, The Chinese University of Hong Kong, Shatin, Hong Kong SAR. }
\thanks{$^{2}$S.L. Charreyron, Q. Boehler, C. Weibel, and B.J Nelson are with the Multi-Scale Robotics Lab, ETH Zurich, Zurich, Switzerland.}%
}

\begin{document}

\maketitle
\thispagestyle{empty}
\pagestyle{empty}

\begin{abstract}
Electromagnetic Navigation Systems (eMNS) can be used to control a variety of multiscale devices within the human body for remote surgery. Accurate modeling of the magnetic fields generated by the electromagnets of an eMNS is crucial for the precise control of these devices. Existing methods assume a linear behavior of these systems, leading to significant modeling errors within nonlinear regions exhibited at higher magnetic fields. In this paper, we use a random forest (RF) and an artificial neural network (ANN) to model the nonlinear behavior of the magnetic fields generated by an eMNS. Both machine learning methods outperformed the state-of-the-art linear multipole electromagnet method (LMEM). The RF and the ANN model reduced the root mean squared error of the LMEM when predicting the field magnitude by around 40\% and 80\%, respectively, over the entire current range of the eMNS. At high current regions, especially between 30 and 35 A, the field-magnitude RMSE improvement of the ANN model over the LMEM was over 35 mT. This study demonstrates the feasibility of using machine learning methods to model an eMNS for medical applications, and its ability to account for complex nonlinear behavior at high currents. The use of machine learning thus shows promise for improving surgical procedures that use magnetic navigation.
\end{abstract}

\section{INTRODUCTION}
Magnetic Navigation Systems (MNS) use magnetic fields to wirelessly control biomedical devices inside the body. These may be untethered magnetic micro or nanorobots that are pulled by magnetostatic forces due to spatially varying magnetic fields \cite{Ullrich2013}, or that ``swim" in fluids due to time-varying magnetic fields \cite{Servant2015}. Additionally, magnetic navigation can be used for steering tethered surgical devices such as ophthalmic microcatheters \cite{Charreyron2018} or endoscopes \cite{Scaglioni2019}. Magnetic navigation has seen clinical adoption for cardiovascular interventions, with MNS systems from Aeon Scientific \cite{Chautems2017} and Stereotaxis Inc. \cite{Ernst2004} achieving clinical certification and performing operations on several thousand patients. Magnetic navigation can also be adopted to control wireless capsules for noninvasive examination of the large gastric cavity \cite{MACE}, and therapeutic functions along the gastrointestinal tract, such as haemostasis \cite{tWCE} and endoscopic submucosal dissection \cite{DBLP:journals/tii/LauLCYLP16}, can potentially be improved with the integration of such systems.

Magnetic navigation can either be performed by sets of moving or rotating permanent magnets \cite{Wright2017}, or by systems comprising several electromagnets \cite{Kummer2010}, also known as Electromagnetic Navigation Systems (eMNS). Modeling a MNS consists of determining the magnetic field flux density at different locations within the workspace, given different varying control parameters such as the permanent magnet placement or the electromagnet currents. By modeling the magnetic fields acting on the steered tools, such as microrobots or catheters, forward kinematic models relating the control variables and the state of the tool can be obtained. The kinematic models can then be inverted to determine the control variables for a desired tool configuration. Therefore, accurately modeling the magnetic fields of a MNS is important for precisely steering the tool. Accurate magnetic models are even more important for tracking devices in a MNS, due to the significant position-dependency of magnetic fields \cite{DiNatali2016, Son2016}. By combining precise measurements of onboard magnetic sensors and an accurate magnetic field map, the pose of a tool in a MNS can be tracked without using line-of-sight, magnetic resonance imaging, or fluoroscopy.


The magnetic vector fields generated by ferromagnets can be modeled using finite-element-method simulation \cite{Sikorski2017}, by interpolating the measured values over space \cite{Ongaro2018}, or using a physics-based multipole model \cite{Petruska2017} that is fit to the measured magnetic field data. When using electromagnets, one must characterize the relationship between the currents applied to the electromagnet coil windings and the resultant magnetic fields. Electromagnets often comprise ferromagnetic cores for magnifying the fields generated by the coils. In previous modeling approaches, the magnetization of such cores are assumed to depend linearly on the magnetic fields that were used to magnetize them. Thus, by the principle of superposition, one can represent the magnetic field flux-density $\mathbf{b} \in \mathbb{R}^3$ generated at a given position $\mathbf{p} \in \mathbb{R}^3$ as the product of an actuation matrix $A \in \mathbb{R}^{3\times N_c}$ and a vector of currents $\mathbf{i} \in \mathbb{R}^{N_c}$, i.e.
\begin{align}
    \mathbf{b}(\mathbf{p}, \mathbf{i}) &= A(\mathbf{p}) ~ \mathbf{i}, \label{eq:linear}
\end{align}
where $\mathbf{i}$ corresponds to the currents in the current windings of the $N_c$ electromagnets.

However, as the external magnetization fields increase, ferromagnetic materials exhibit saturation, and the relationship between coil currents and the generated magnetic field is no-longer linear. Due to these nonlinearities, the superposition-principle does not hold, and it is impossible to separate the effect of different coils. A more general expression $g$ for the magnetic fields must be used to take into account the effects of all coil currents, i.e.
\begin{align}
    \mathbf{b}(\mathbf{p}, \mathbf{i}) &= g(\mathbf{p}, \mathbf{i}). \label{eq:nonlinear}
\end{align}

For systems with a small number of electromagnets, $g$ can be determined by measuring discrete magnetic fields that span the entire space of currents, and then interpolating a smooth function between the field measurements. However, for systems consisting of more than three electromagnets, such an approach becomes computationally infeasible due to the ``curse of dimensionality". 

In this work, we propose two machine learning methods to model the magnetic fields generated by the CardioMag, an eMNS exhibiting strong saturation over 70\% of its magnetic field generation capacity. For comparison, a state-of-the-art linear model is used as the baseline.

This paper is organized as follows. We first introduce the baseline model and the applied machine learning models in \ref{sec:models}. We subsequently detail the data collection and model training processes in \ref{sec:experiments}. We present the obtained results in \ref{sec:results} followed by a discussion in  \ref{sec:discussion}, and conclude in \ref{sec:conclusion}.

\section{MODELING METHODS}
\label{sec:models}

Two supervised machine learning methods, a random forest (RF) and an artificial neural network (ANN), are used to model an eMNS. Both approaches are compared to a linear multipole electromagnetic method (LMEM) introduced in the literature, which constitutes our baseline.  

\subsection{Linear Multipole Electromagnet Method (Baseline)}

The LMEM uses a multi-source spherical multipole expansion to describe the magnetic scalar potential produced by a set of electromagnets with ferromagnetic cores \cite{Petruska2017}. This formulation ensures that the magnetic field associated with this scalar potential is curl-free and divergence-free, which constitute two fundamental physical properties of the field. This is because the multipole expansion is a solution to Laplace's equation, which defines the magnetic scalar potential. This method has been previously used to model the CardioMag.

The model assumes a linear relationship between the magnetic fields and the coil currents, and superimposes the contribution of each electromagnet to predict the magnetic field. It neither considers the nonlinearities that occur within the saturation region of the ferromagnetic cores, nor the perturbations in the magnetic field resulting from other unidentified sources.

\subsection{Machine Learning Methods}
In this work, we use data collected from magnetic flux density sensors placed over the workspace of the CardioMag, at a number of electromagnet currents, to train both machine learning models. The task at hand is to predict the generated 3-D magnetic field flux density at a specific position, given the electrical currents that are measured on the electromagnets. The prediction output is a vector that contains the continuous 3-D magnetic field flux density. Since multivariate regression is needed to predict these three values describing the field, a RF and an ANN were used in this study.

A RF \cite{Breiman2001} is an ensemble learning method, where predictions are made by growing multiple decision trees. Each tree performs a binary split of samples at each node by considering a subset of features. For regression problems, final predictions are made in a RF by averaging the results of all trees. An ANN \cite{Mitchell:1997:ML:541177} contains many connected neurons arranged in layers to produce network outputs. ANNs are motivated by biological neural systems and can be trained to minimize the error between the network output values and the target values using the backpropagation algorithm.

\section{EXPERIMENTS}
\label{sec:experiments}

In this section, we introduce the data collection process, including the hardware setup as well as the data collection protocol, and the model training and the evaluation process.

\subsection{Hardware Setup} 

\begin{figure}[tpb]
    \centering
    \includegraphics[scale=0.7]{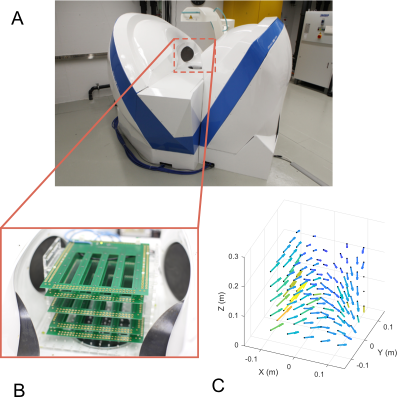}
    \caption{Data collection setup for an eMNS. A: The CardioMag, an eMNS B: Magnetic sensor array C: Magnetic field measurements with the sensor array for a random current set}
    \label{cmag}
\end{figure}

The eMNS to be modeled was introduced in~\cite{Petruska2017} and is depicted in Fig.~\ref{cmag}.A. It comprises $N_c = 8$ electromagnets surrounding a $10 \times 10 \times 10$ \SI{}{\centi\meter} cubic workspace. The maximum current in each electromagnet is \SI{35}{\ampere}, and the maximum power in the whole system is \SI{15}{\kilo\watt}.

To obtain 3-D magnetic field measurements within the workspace, an array of magnetic sensors was built, as shown in Fig. \ref{cmag}.B. The array consists of 125 identical Hall-effect magnetic sensors (TLV493D-A1B6, Infineon) \cite{infineon:TLE4473GV55-2}, arranged in a $5 \times 5 \times 5$ cubic grid with 5 cm spacing in each direction.



\subsection{Data Collection} 

A set of uniformly random current vectors with values between -\SI{35}{\ampere} and \SI{35}{\ampere} was first generated. Current vectors exceeding the maximum system power were discarded. A total number of 3,590 distinct current vectors were generated for the dataset. The sensor array was placed in the center of the workspace, as shown in Fig.~\ref{cmag}.B. The pre-generated current vectors were applied to the system and the resultant magnetic fields were recorded by the sensors at a frequency of \SI{1}{\hertz}.

The raw measurements from the magnetic sensors were then preprocessed to construct a dataset for experiments. Since the electromagnets exhibit a dynamic response, the transient region of measurements was discarded to obtain static measurements only. The currents that were measured on the coil windings had insignificant white noise, with a mean standard deviation of \SI{169}{\milli\ampere}. Due to the slow dynamic response of the coils, the effect of such high-frequency noise had little effect on the generated magnetic fields. Nonetheless, the current measurements were smoothed by averaging their values over the measurement window. The mean standard deviation of the magnetic field measurement was \SI{148}{\micro\tesla}. Magnetic field measurements were also averaged to reduce the effect of such measurement noise.

The dataset consisted of $M = $~427,210 samples obtained from 119 sensors\footnote{Six sensors were malfunctioning during the data collection process, and their measurements were thus removed from the dataset.}. Table~\ref{tbl_dataset} shows the statistics of the dataset. Each recorded sample $j \in [1,M]$ consisted of: 1) a position vector $\mathbf{p}^j = \begin{bmatrix}x^j & y^j & z^j \end{bmatrix}^T$ of the sensor; 2) a current vector $\mathbf{i}^j = \begin{bmatrix}i^j_1 & \dots & i^j_8 \end{bmatrix}^T$, where $i^j_k$ indicates the smoothed current applied to the $k$-th coil with $k \in [1,8]$; 3) a magnetic field vector $\mathbf{b}^j = \begin{bmatrix}b^j_x & b^j_y & b^j_z \end{bmatrix}^T$ measured at $\mathbf{p}^j$. The magnitude of the magnetic field corresponds to the magnetic flux density magnitude $\|\mathbf{b}^j\|$. A sample extracted from the dataset is depicted in Fig.~\ref{cmag}.C, where each arrow represents the measured magnetic field at a sensor position in space.

\begin{table}[h]
\caption{Statistical information about the dataset.}
\label{tbl_dataset}
\begin{center}
\begin{tabular}{@{}llll@{}}
\toprule
\textbf{Parameter} & \textbf{Minimum} & \textbf{Maximum} & \textbf{Unit}\\ \midrule
x & -10.12 & 10.12 & cm\\
y & -10.61 & 12.33 & cm\\
z & 1.94 & 22.20 & cm\\
$i_k$ for $k \in [1,8]$ & -35.00 & 34.99 & A\\
$b_x$ & -179.35 & 178.46 & mT\\
$b_y$ & -166.81 & 170.45 & mT\\
$b_z$ & -179.50 & 183.89 & mT\\ \bottomrule
\end{tabular}
\end{center}
\end{table}

\subsection{Model Training}
The collected current vectors were randomly divided into a training and testing dataset with a 9:1 ratio. For all models, the input data consisted of an 11-dimensional vector concatenating the position in the workspace $\mathbf{p} \in \mathbb{R}^3$ and the electromagnet current vector $\mathbf{i} \in \mathbb{R}^8$. Each model output a 3-D magnetic field $\mathbf{b} \in \mathbb{R}^3$.
To generate a training dataset for the LMEM, we followed the original requirements in \cite{Petruska2017} where the maximum current in each coil was limited to 5 A. As neural networks are sensitive to the scale of inputs, all features ($\textbf{p}$ and $\textbf{i}$) were scaled between 0 and 1 using the min-max scaling method based on the statistics calculated from the training dataset similar to those shown in Table \ref{tbl_dataset}.

The RF model was implemented using the scikit-learn package \cite{Pedregosa2011}. A five-fold cross-validation grid search was performed to select hyperparameters for the model. The searched parameter grid covered the number of trees between 10 and 100, the maximum depth of each tree between 10 and 30, the minimum number of samples to split at each node between 2 and 20, the maximum number of features to consider at each node between 3 and 5, and the minimum number of samples at a leaf node between 1 and 15. The best performing model had 100 trees with a maximum depth of 25, and a maximum of 5 features to consider. It required at least 2 samples at each node and 1 sample at each leaf node. 

The ANN model was implemented using the Keras library \cite{CholletFrancois2015}. The model structure was adopted from a study \cite{Christensen2017} with similar feature dimensions. With 11 neurons in the input layer, the implemented ANN contained three hidden layers with 100, 50 and 25 neurons, respectively. The hyperbolic tangent function (tanh) was selected as the activation function in each hidden layer. Finally, the output layer had three neurons with a linear activation function. During training, 10\% of training data were set aside for validation. The ANN model was trained using the Adam \cite{adam} optimizer with an initial learning rate of 0.001, in order to minimize a mean squared error between the predicted and the measured magnetic fields. The batch size was chosen as 128 samples, and the epoch number was 50. To prevent overfitting, early stopping was applied during training when the validation loss did not decrease in 5 consecutive epochs. The model with the lowest validation loss was selected for testing. 

The size of the training data is an important factor limiting the performance of machine learning models. In this study, we conduct further experiments to evaluate the impact of the size of training data on the prediction accuracy for the RF and ANN models. Volume of 10\% to 90\% of the training samples was randomly selected from the original training dataset to construct multiple smaller training subsets. The RF and ANN models were independently trained on these training subsets. We used the same model hyperparameters and training process as described previously. All trained models were then tested on the original testing dataset for comparison. 

\subsection{Evaluation Metrics}

To evaluate the prediction performance of the models, two general goodness-of-fit metrics were used to compare the measured and predicted magnetic field. These included the $R^2$-score and the root mean squared error (RMSE) for each component computed as

\begin{align} R_\star^2 = 1 - \frac{\sum_{j=1}^{N} ({{b}}^j_{\star}- {\hat{b}^j}_\star)^2}{\sum_{j=1}^{N} ({b}^j_{\star} - \bar{{b}}_\star)^2},\end{align}

and 
\begin{align}
\text{RMSE}_{\star} = \sqrt{\frac{\sum_{j=1}^{N}(b^j_{\star} - \hat{b}^j_{\star})^2}{N}},
\end{align}
where $b^j_\star$ and $\hat{b}^j_\star$ are respectively the measured and model predicted values for the $j$-th sample and the $\star$ component; $\bar{b}_\star$ is the mean of the measured magnetic field over the $N$ samples composing the testing dataset. Additionally, the prediction performance on the magnetic field magnitude was also evaluated using these two metrics denoted as $R^2_\text{norm}$ and $\text{RMSE}_\text{norm}$. An $R^2$ value of 1 indicates that the model predictions perfectly fit the measurements, whereas a RMSE close to 0 suggests a good model.

To evaluate the models' prediction performance at different locations, the mean absolute percentage error of the magnetic field magnitude at location \textbf{p} is calculated by 
\begin{align}
\text{MAPE}_{norm}^\textbf{p} = \frac{100\%}{K} \sum_{k=1}^{K} \left |\frac{ \|\mathbf{b}^\mathbf{p}_k\| - \|\hat{\mathbf{b}}^\mathbf{p}_k\| }{\|\mathbf{b}^\mathbf{p}_k\|}\right |,
\end{align}
where $\|\textbf{b}^\textbf{p}_k\|$ and $\|\hat{\textbf{b}}^\textbf{p}_k\|$ are respectively measured and predicted magnetic flux density magnitude at location \textbf{p}, and k is the index of the current vector with a total K currents vectors tested at each location.

\section{RESULTS}
\label{sec:results}
The overall testing performance of the LMEM, RF, and ANN models are summarized in Table \ref{tbl_overall_test}. The LMEM achieved over 0.75 $R^2$ for all field components, while only 0.29 $R^2$ for the field magnitude. The LMEM produced at least \SI{14}{\milli\tesla} component-wise RMSE, and collectively nearly \SI{24}{\milli\tesla} field-magnitude RMSE. The RF and ANN models achieved significantly better results. The RF model achieved over 0.85 $R^2$ in all field components and 0.74 in field-magnitude $R^2$. The RF model produced around 30\% improvement over the baseline model based on the component-wise RMSE and a 40\% improvement on the field-magnitude RMSE. The ANN model achieved 0.99 $R^2$ in predicting each field component and magnitude. The ANN model showed an 80\% improvement over the LMEM based on both the component-wise and field-magnitude RMSE. 

\begin{table*}[!htbp]
\ra{1.3}
\caption{Performance comparison of the three models.}
\label{tbl_overall_test}
\centering
\begin{tabular}{@{}lllllllll@{}}
\toprule
\textbf{Model} & $\mathbf{R^2_x}$ & $\mathbf{R^2_y}$ & $\mathbf{R^2_z}$ & $\mathbf{R^2_{norm}}$ & $\mathbf{RMSE_x (mT)} $ & $\mathbf{RMSE_y (mT)} $ & $\mathbf{RMSE_z (mT)} $ & $\mathbf{RMSE_{norm} (mT)} $\\ \midrule
LMEM & 0.86 & 0.81 & 0.76 & 0.29 & 14.34 & 15.81 & 14.51 & 23.90\\
RF & 0.92 & 0.92 & 0.86 & 0.74 & 10.89 & 10.00 & 11.11 & 14.43\\
ANN & 0.99 & 0.99 & 0.99 & 0.99 & 3.10 & 2.68 & 2.72 & 3.01\\ \bottomrule
\end{tabular}
\end{table*}



\begin{figure}[thpb]
    \centering
    \includegraphics[scale=0.45]{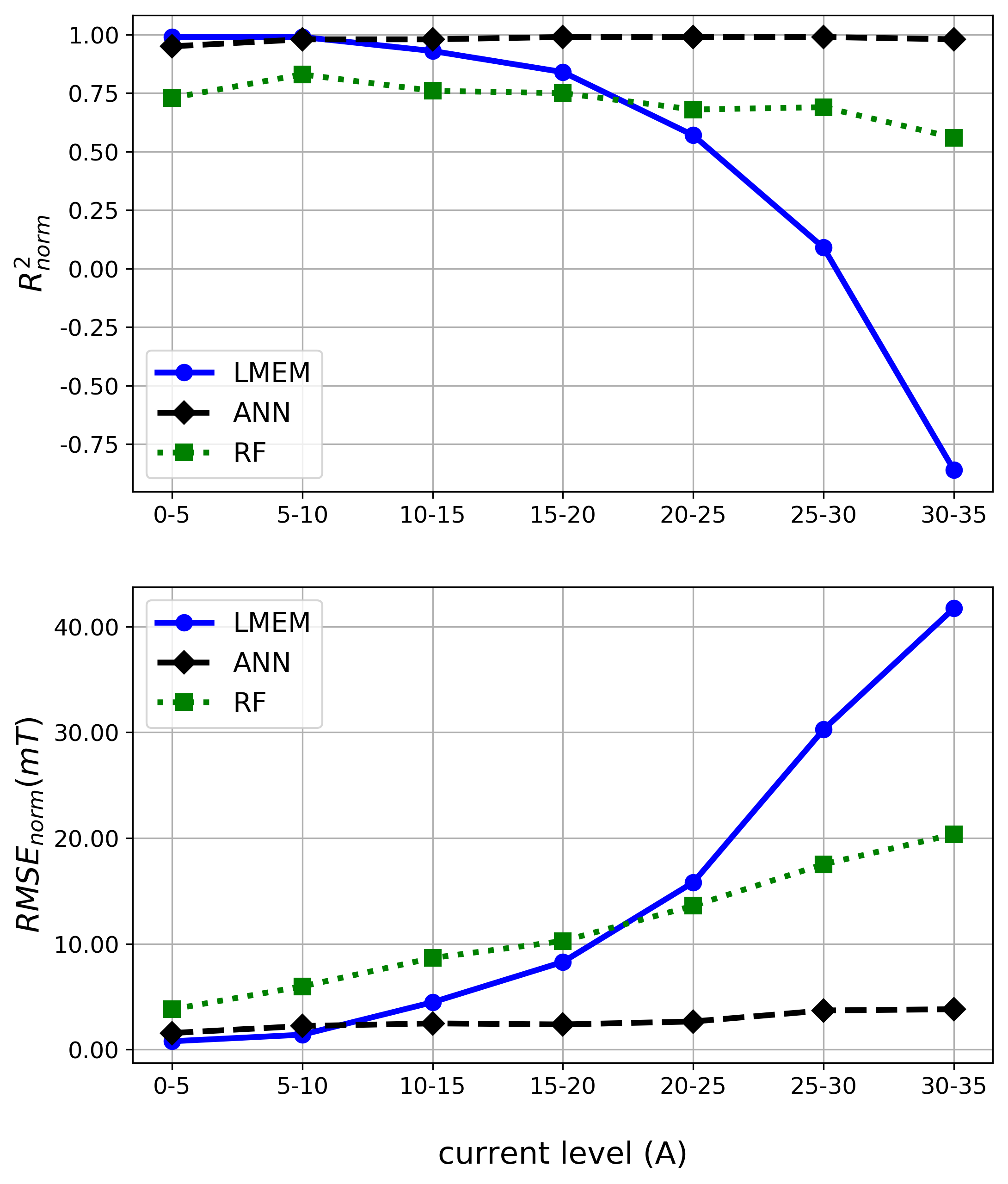}
    \caption{Prediction performance comparison in the testing dataset stratified by current levels.}
    \label{fig_Results_diff_amps}
\end{figure}

To further examine the prediction models' performance at different currents, testing samples were grouped into different current levels according to the maximum electromagnet current in the current vector $i^j_{\max} = \max(|i^j_1|, \dots ,|i^j_8|)$. Predictions of the generated field magnitudes were then evaluated independently for different current levels, as shown in Fig. \ref{fig_Results_diff_amps}. For both metrics, the performance of the LMEM had a tendency to decrease as the current level increased. The RF and ANN models had relatively stable performance across all current levels in terms of $R^2_\text{norm}$. The RF model also showed an increasing $\text{RMSE}_\text{norm}$ as currents increased. The ANN model showed better performance than the RF model across all current levels. When applied currents were small, where linear assumptions of the LMEM still held, the LMEM and the ANN model showed similar performance, while the RF performed worse. For testing samples with the maximum current over \SI{10}{\ampere}, the ANN model performed better than the LMEM. The superior performance of the RF model over the LMEM was shown when the maximum current exceeded \SI{20}{\ampere}. When the maximum applied current was within 30-35 \SI{}{\ampere}, the improvement of the RF and ANN models over the LMEM were \SI{20}{\milli\tesla} and \SI{35}{\milli\tesla} respectively in terms of field-magnitude RMSE.


\begin{figure*}[thpb]
    \centering
    \includegraphics[scale=0.5]{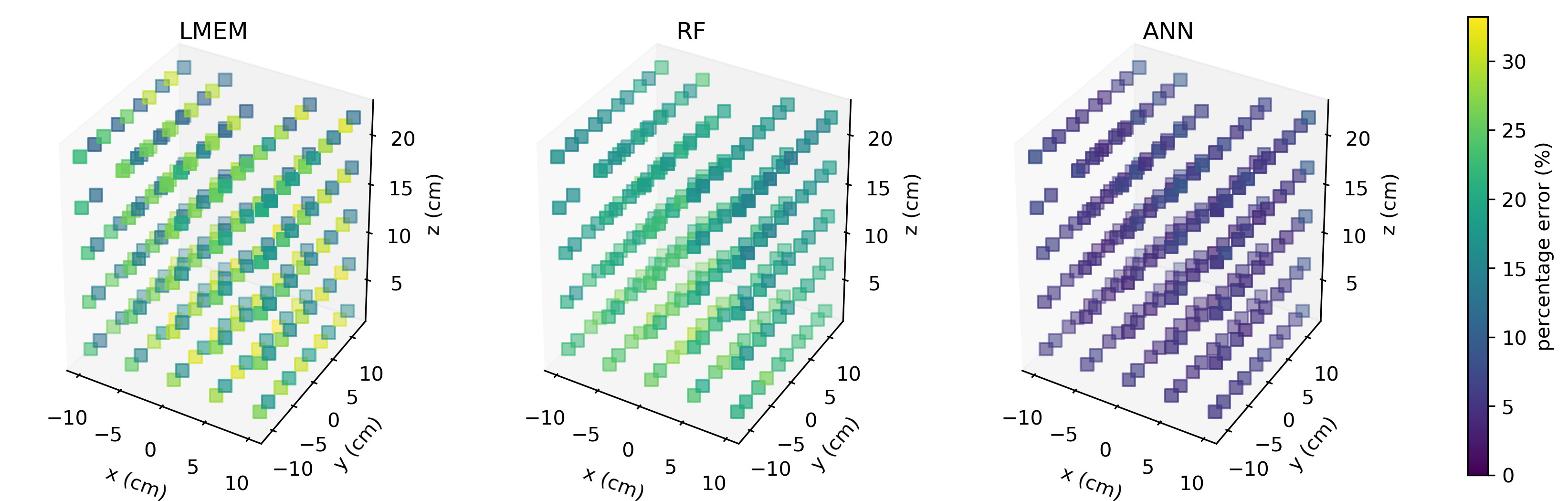}
    \caption{Spatial prediction error comparison among three models. The maximum $\text{MAPE}_{norm}^\textbf{p}$ across all sensor locations was 33.24\% for the LMEM, 29.38\% for the RF model and 10.91\% for the ANN model.}
    \label{spatial_error}
\end{figure*}

To examine the spatial modeling error distribution, predictions of the three models were evaluated at all sensor locations. The $\text{MAPE}_{norm}$ was calculated at each location as depicted in Fig.~\ref{spatial_error} for all samples in the testing dataset. Both LMEM and RF produced significantly higher $\text{MAPE}_{norm}$ than the ANN model at all evaluated locations. The RF model showed slightly better prediction performance than the LMEM.

\begin{figure}[thpb]
    \centering
    \includegraphics[scale=0.4]{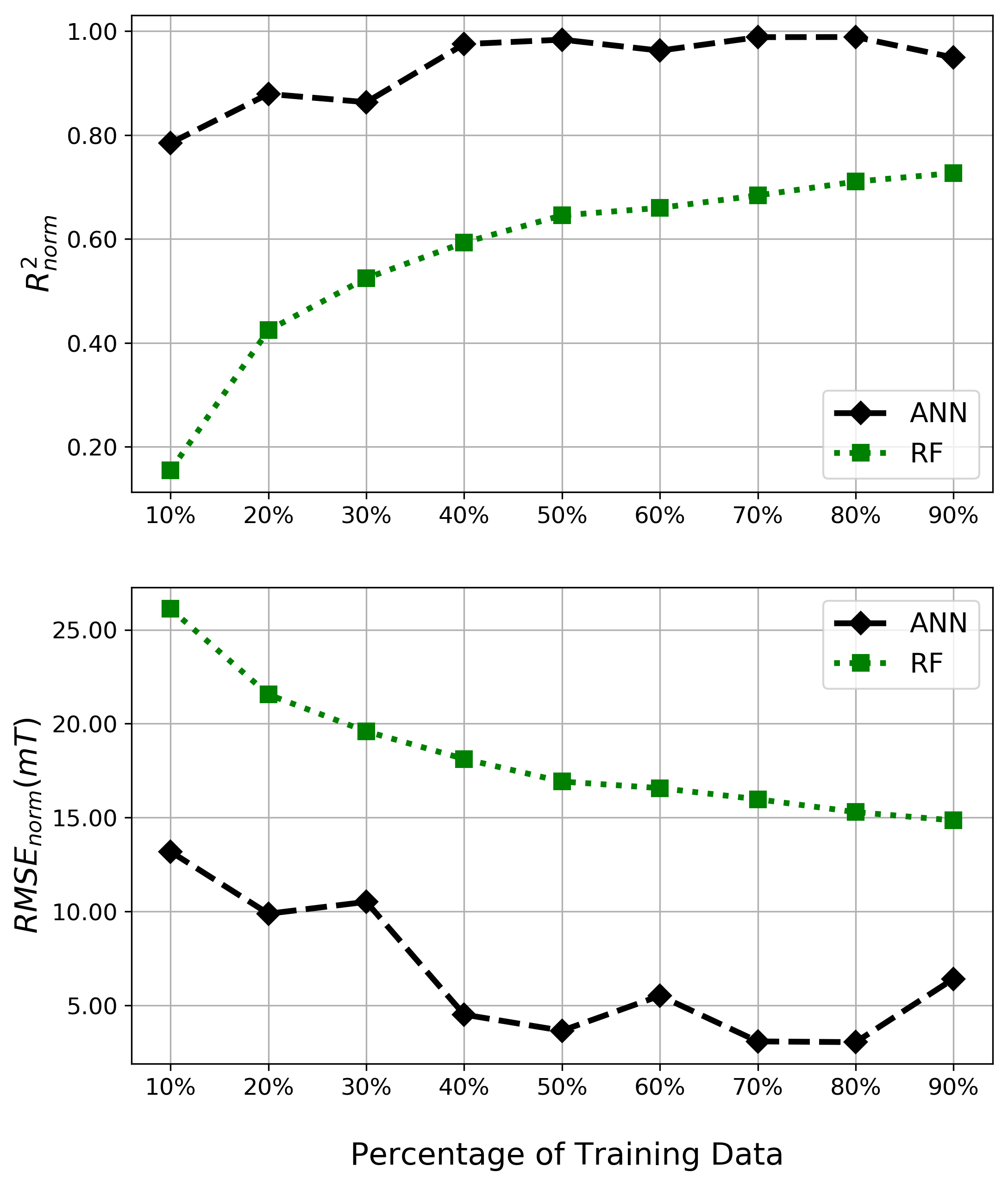}
    \caption{The impact of the training set size on prediction performance.}
    \label{fig_train_size}
\end{figure}

Fig. \ref{fig_train_size} shows the results of the testing performance of the RF and ANN models when trained with different amounts of training data. In general, both machine learning methods showed an increase in performance with the increasing amount of training data. Compared with the ANN model, the RF model's performance exhibited a more significant performance improvement when supplied with more training data. For both models, the performance gain started to decrease when training subsets were over 40\% of the original training data, especially for the ANN model.


\section{DISCUSSION}
\label{sec:discussion}


eMNS can be designed for specific surgical applications, and different numbers, or different configurations of electromagnets can be used to maximize the resultant magnetic fields within a sufficiently large workspace for the operation. Modeling a given eMNS can be carried out prior to deployment with a protocol similar to the one described in this study. Our modeling approach can be used to model any eMNS regardless of  the workspace size as well as the number and properties of the electromagnets.

Overall, both implemented machine learning models performed better than the LMEM on the entire testing dataset. The LMEM was able to model the magnetic fields precisely at low currents but not at currents higher than 15-20 A. This was as expected, since the linear assumption does not hold at these currents. The development of the LMEM requires prior knowledge on the geometry and strength of the dipoles which model the ferromagnetic sources of the eMNS, which in some cases are difficult to define. Although samples consisted of a wide range of current levels and spatial locations, relatively stable prediction performance was achieved by both machine learning models, especially the ANN model. 

The ANN model performed better than the RF model for all evaluation metrics in all scenarios. Since the regression output from a RF model is predicted by averaging results from all trees, and the prediction of each tree depends on the samples arrived at the leaf node, only a finite number of potential prediction outputs are possible once trained. When used in reality, additional steps are required to interpolate the RF field predictions between locations and current vectors. The ANN model, on the other hand, can directly output continuous prediction values within the range of the activation function in the output layer. In this case, the ANN model may be a better prediction method to model the continuous magnetic fields generated by an eMNS. Although RF was not the most precise method for modeling the eMNS, it could compute the relative importance of features on predicting the magnetic fields. The higher the value, the more important the feature is in the prediction. The feature importance values returned from our RF model were as follows: $i_8$ (0.15), $x$ (0.12), $i_2$ (0.12), $i_4$ (0.11), $i_6$ (0.11), $y$ (0.09), $z$ (0.08), $i_1$ (0.06), $i_7$ (0.06), $i_3$ (0.05), $i_5$ (0.05).  All features contributed on a similar level of importance to the magnetic field prediction. However, from the final RF model's perspective, a location's coordinate along the x-direction was slightly more important than the other two directions. Moreover, currents of the even-numbered coils were, in general, more important than those of the odd-numbered coils from these returned feature importance values. These values provide insights into understanding the MNS behavior for those who are unfamiliar with the system. In addition, when considering additional factors which may relate to the magnetic field prediction, these values can be used for selecting important predictors for modeling approaches.

The sample size of the training data is critical for both RF and ANN models to achieve good performance. In this study, we evaluated the influence of the size of the training data on model performance. As anticipated, the performance of both RF and ANN models improved when supplying the model with more training data. Since the RF model cannot extrapolate target values, increasing the training size may potentially increase the range of values it can predict, and hence leading to better performance.    

The target modeling performance will depend on the specific application of the eMNS, and whether the modeled magnetic field map is going to be used to determine control variables applied to the system or the states of the devices. In general, there is no ceiling on the performance improvement, but some potential applications like localization would require performance that is much higher than what is achieved by the LMEM.

\section{CONCLUSIONS}
\label{sec:conclusion}

We presented two machine learning approaches to model a medical eMNS, namely using RF and ANN models. The results of both machine learning models were compared to a state-of-the-art LMEM. Both RF and ANN models achieved better overall performance than the LMEM in terms of $R^2$ and RMSE. The ANN model achieved even better modeling performance than the RF model, as an improvement over the LMEM was over 80\% as opposed to 40\% in terms of field-magnitude RMSE. The RMSE improvement of the ANN model over the LMEM model exceeded \SI{35}{\milli\tesla} when the applied current was in the range of 30-35 A, while the improvement of the RF model was around \SI{20}{\milli\tesla}. Machine learning shows promise for improving the precision of surgical procedures that use magnetic navigation by improving the magnetic field prediction.


\section*{ACKNOWLEDGMENT}
This work was done when R. Yu was visiting the Multi-Scale Robotics Lab, ETH Zurich. The research activity was supported in part by the CUHK Research Postgraduate Student Grants for Overseas Academic Activities, Hong Kong Innovation and Technology Fund, and General Research Fund.

We would like to thank NVIDIA for providing us with a Titan Xp through the GPU grant program. This work was also supported by the Swiss National Science Foundation through grant number 200021 165564.

\balance
\bibliographystyle{IEEEtran}
\bibliography{references}


\end{document}